\documentclass[letter,
               keeplastbox,   
               ]{jacow}
\usepackage{pdfpages,multirow,ragged2e,appendix} %
\makeatletter%
	\ifboolexpr{bool{xetex}}
	 {\renewcommand{\Gin@extensions}{.pdf,%
	                    .png,.jpg,.bmp,.pict,.tif,.psd,.mac,.sga,.tga,.gif,%
	                    .eps,.ps,%
	                    }}{}
\makeatother

\ifboolexpr{bool{xetex} or bool{luatex}} 
 {}                                      
 {\usepackage[utf8]{inputenc}}           

\usepackage[USenglish]{babel}

\ifboolexpr{bool{jacowbiblatex}}%
 {%
  \addbibresource{jacow-test.bib}
  \addbibresource{biblatex-examples.bib}
 }{}
\begin{document}

\title{Synchronization and Phase Locking of Resonant Magnet Power Supplies for \NoCaseChange{Mu2e} Experiment at Fermilab\thanks{This work is supported by US Department of Energy grant DE-SC0009999.}}

\author{R. Hensley\thanks{rshensley@ucdavis.edu}, E. Prebys, S. Tripathy\\ University of California Davis, Davis, CA, USA}
	
\maketitle

\begin{abstract}
    The Muon-to-Electron Conversion (Mu2e) Experiment demands a highly precise magnet and collimator system to achieve a stringent extinction level of $1\times 10^{-10}$ for out-of-time beam particles. Extinction is ensured by an AC Dipole system consisting of two magnet components: a 295 kHz system to allow for the passage of a 590 kHz beam at the nodes, and a 4.42 MHz system to minimize in-time beam slewing. Both components must be accurately phase-locked to the Delivery Ring's bunch rate as well as be synchronized with beam transfers from the Recycler.
    
    In this paper, we present the design, implementation, and results of a control system for the Mu2e magnet system based on an Intel Arria 10 FPGA. This system handles the phase-locking of the magnets to the Delivery Ring, as well as the phase jumps required for synchronization with transfers from the Recycler.
    
\end{abstract}






\section{Background}

The Mu2e experiment is designed to search for the direct conversion of a muon captured in the orbit of a nucleus to an electron without the production of any neutrinos ($\mu N \rightarrow~e N$)~\cite{TDR}. Observing this process would show charged lepton flavor number violation and be direct, unambiguous evidence of physics beyond the Standard Model.

\begin{figure}[!htb]
   \centering
   \includegraphics*[width=\columnwidth]{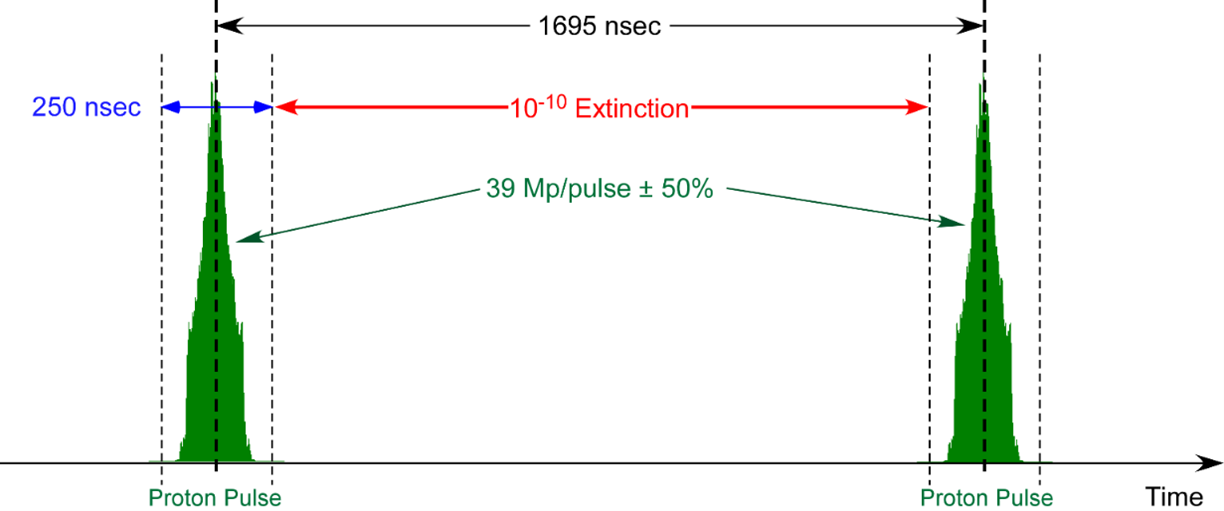}
   \caption{Pulsed proton beam microstructure. Proton pulses are separated by 1,695 nsec and must have a level of $10^{-10}$ extinction between pulses.}
   \label{fig:proton_pulse}
\end{figure}

The longitudinal timing and structure of the pulsed proton beam used in this experiment is crucial to the success of the Mu2e experiment\cite{TDR}. During the experiment, the muons for the experiment will be produced by proton pulses incident on a tungsten target creating negative pions, and the proton pulses must each be separated in time by 1,695 nsec as is shown in Fig.~\ref{fig:proton_pulse}. Each pulse will contain protons within a transmission window of $\pm 125$ nsec. Particles within this window are called "in-time" protons, and if any particles make it to the proton target outside of the transmission window, they are called "out-of-time" protons. To minimize background, it is important to reduce the ratio of the number of out-of-time protons to in-time protons as much as possible, which is a condition called extinction. The requirement for extinction in the Mu2e experiment is a relative extinction level of $10^{-10}$\cite{extinction}.

\section{AC Dipole Beam Extinction System}

An AC Dipole is going to be the biggest source of extinction in this experiment, and it will be controlled by the FPGA discussed in the latter half of this paper. It will be surrounded upstream and downstream by multiple collimators, and is designed such that the particle beam should only make it through this combination of collimators and the AC Dipole when there is no field in the AC Dipole magnets\cite{TDR, Werkema}. If there is no field, the proton pulse will proceed through all the collimators without any interference. However, if there is out-of-time beam that comes through while the AC Dipole field is active, those particles will be pushed into the extinction collimator located just downstream of the AC Dipole. As a result, only the beam that comes through the AC Dipole in the nodes of its magnetic field make it through, and extinction occurs everywhere else.

Before the AC Dipole, an extinction level of $10^{-5}$ is achieved naturally using accelerator dynamics, and then the AC Dipole will add another extinction factor of below $10^{-7}$. Combining these two factors, the extinction level of $10^{-10}$ required by the experiment should be achievable. Tests are currently underway to measure the extinction level out of the Delivery Ring and before the AC Dipole to see if an extinction level of $10^{-5}$ is being achieved as hoped from the first stage. 

\begin{figure*}[!tbh]
    \centering
    \includegraphics*[width=\textwidth]{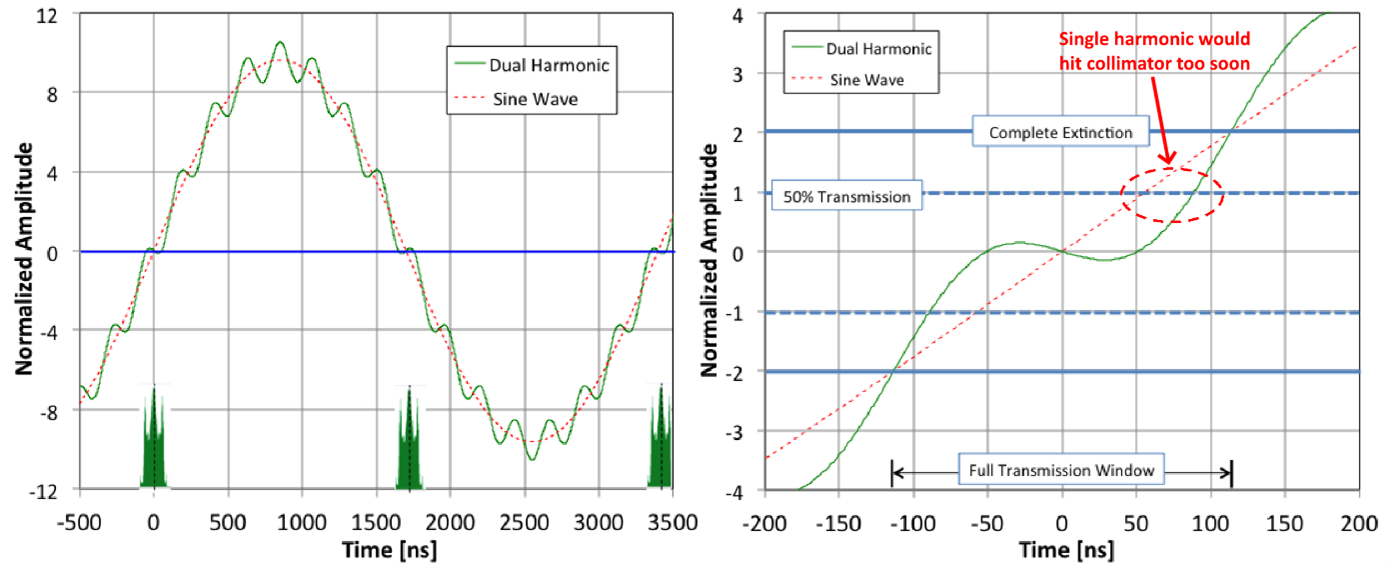}

    \caption{A plot of the dual harmonic system of the AC Dipole that the FPGA code must create. The 295 kHz wave is the main body of the waveform, and the 15th harmonic reduces in-time beam slewing and enlarges the transmission window slightly.}
    \label{fig:power_supply}
\end{figure*}

\begin{figure*}[!tbh]
    \centering
    \includegraphics*[width=\textwidth]{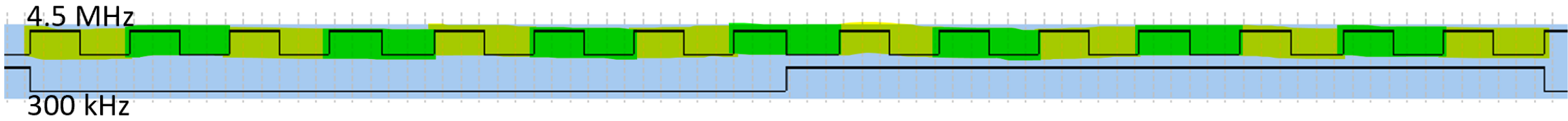}

    \caption{The FPGA output showing 15 periods of the 4.5 (4.42) MHz wave per one period of the 300 (295) kHz wave. Note that the 300 kHz wave transitions on the falling edge of a 4.5 MHz wave period.\vspace{-2mm}}
    \label{fig:fpga_output}
\end{figure*}

The AC Dipole will be powered by a combination of 295 kHz and 4.42 MHz power supplies%
\footnote{In some documentation, these frequencies are seen rounded as 300 kHz and 4.5 MHz. The exact number comes from the requirement for a half-period of 1,695 nsec in the 295 kHz wave which gives 294.985 kHz, and then 4.42 MHz is the 15th harmonic of that.}%
and must switch at precise times to ensure maximum extinction. The 295 kHz frequency is chosen to match the rate of incoming proton pulses from the Delivery Ring so they come in at the nodes of the magnetic field, and the 15th harmonic 4.42 MHz wave is placed as modulation on top of the carrier 295 kHz wave to reduce the slewing of in-time beam. The result of this modulation is shown in Fig.~\ref{fig:power_supply}, and the 15th harmonic alignment with the carrier wave is highlighted in Fig.~\ref{fig:fpga_output}. 

There are multiple times where there will be an interruption in the incoming pulsed beam of protons, so phase jumps will be necessary whenever the pulsed beam returns after one of these interruptions.

Physically, the AC Dipole will consist of a series of six magnets in two vessels, with each magnet being powered by a separate power supply\cite{ac_dipole}. The 295 kHz module will be powered by a slightly modified version of a switching power supply that has been frequently used at Fermilab. The 4.42~MHz module will be powered by a commercial RF power supply. The power supplies will be controlled by Intel Arria 10 SoC SoM FPGAs mounted on custom Fermilab boards.

\section{FPGA Controls and PLL}

To create the required 295 kHz and 4.42 MHz frequencies, an FPGA with a phase-locked loop (PLL) implemented is a common tool. Most modern FPGA boards have a PLL block implemented with analog circuit elements through the IP catalog in the design program which can create a perfect output frequency with little or no error for whatever frequency the programmer desires.

However, the disadvantage to using one of these is that one has to remake the design each time they switch to a different board. For this experiment, the analog-driven IP catalog version of the PLL was not used, and instead a custom PLL using only digital logic was designed. 

Using only digital logic has a serious limitation in that it means the output clocks can only change polarity on edges of the input clock, but with a desired output half-period of 1,695 nsec, this is actually manageable and does not cause a problem. As a result, the design is based on only pure VHDL code that can be used on any modern FPGA chip from any maker. This code can be viewed from the site at reference \cite{github}.

\subsection{4.42 MHz Clock}

Given an input clock of 100 MHz (10 nsec period), the first task is to make the 4.42 MHz (226 nsec period, 113~nsec per transition) wave. Working with the limit that output clock polarity shifts can only happen on edges of the input clock cycle, one could wait for 11 cycles of the input clock cycle taking 110 nsec, but there is no way to shift three nanoseconds later at 113 nsec for the first required transition. 

Therefore, the closest that one can do is to transition either early after 110 nsec or late after 115 nsec%
\footnote{For the transitions after 115 nsec, special logic is used to transition on falling edges on the input clock.}%
. In this case, the FPGA has been set up to find whether it is ideal to transition after either 110 nsec or 115 nsec. In the end, the FPGA follows a sequence of output clock cycles which each take a number of input clock cycles equal to 22, 23, 22, 23, 23. In total, these five output cycles take 113 input cycles (each being 10 nsec) to complete for a total of 1,130 nsec, but more importantly, $\textit{an average of 226 nsec per cycle}$%
\footnote{Specifically, the design is a phase accumulator that increments a 12-bit counter by 181 every input clock cycle (approximately 22.6 increments to overflow the counter). After each five-output-period supercycle containing 113 input cycles, the phase counter resets to zero to account for the accumulation of error.}%
.

The final result is an output clock that averages 4.42 MHz over five cycles, but what is important is that there it is setup to transition every 1,695 nsec, so the requirement for the creation of the pulsed beam is fulfilled.

\subsection{295 kHz Clock}

Once the 4.42 MHz clock is made, generating a 295 kHz clock (3,390 nsec period) is easy because the 4.42 MHz clock is simply the 15th harmonic of the 295 kHz clock. Therefore, this clock can be generated by counting 7.5 clock cycles (1,695 nsec) of the 4.42 MHz clock and making a transition for the output of a 295 kHz wave. As before, one transition will need to be on the falling edge of the input clock (the 4.42 MHz clock in this case, see Fig.~\ref{fig:fpga_output}). This manner of using the generated output clock from a previous circuit as the input to a new clock generation circuit is called clock cascading.

\subsection{Required Phase Jumps}

\begin{figure}[!tb]
    \centering
    \includegraphics*[width=\columnwidth]{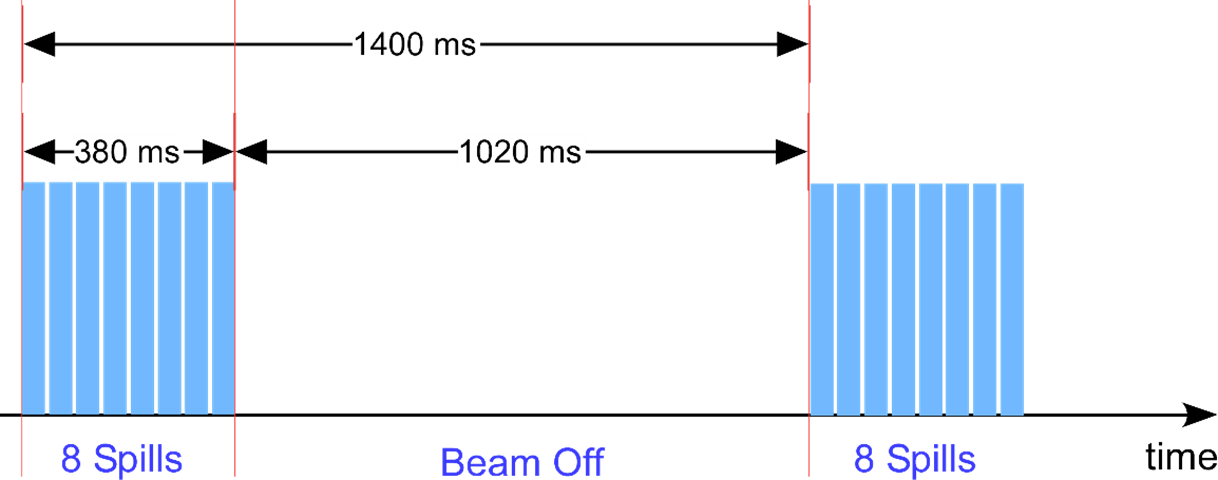}

    \caption{Macro-structure of the proton beam through the AC Dipole and onto the target. There are 5 msec gaps between each of the eight spills during the 380 msec of active beam time, and a long 1,020 msec gap where the beam to the Mu2e experiment is off. In each of these cases, phase jumps will be required in the AC Dipole power supply systems to make sure they stay in phase with the incoming beam.}
    \label{fig:beam_macrostructure}
\end{figure}

There will be a few times where phase jumps are required in the control of the power supplies. First of all, during the active experiment time, proton pulses will come in intervals of 1,695 nsec, but that will only continue for 380 msec, after which the beam will be down for 1,020 nsec while the facilities are being used by the NO$\nu$A experiment. Furthermore, inside those 380 msec will be small gaps of 5 msec of no beam to allow the Delivery Ring to reset its RF systems. This timeline is shown in Fig.~\ref{fig:beam_macrostructure}. Phase jumps in the FPGA control system will be required wherever there is a gap in Fig.~\ref{fig:beam_macrostructure}.

\section{Final Product and Conclusion}

In the end, a circuit was made that can transition very reliably following a 295 kHz waveform every 1,695 nsec (total output period is 3,390 nsec) with a precise 4.42 MHz modulation on top of it to reduce in-time beam slewing. Because of the issue with not being able to transition exactly every 113 nsec to make a perfect 4.42 MHz wave, the duty cycle of the generated wave is not exactly 50\%%
\footnote{Inside each 3,390 nsec cycle of 4.42 MHz waves is nine 230 nsec cycles and six 220 nsec cycles}%
. However, all that is important is that when the pulsed beam comes in every 1,695 nsec, both the 4.42 MHz wave and the 295~kHz wave make a timely transition at the center of the beam transmission window, which is what is accomplished. It is not as important what the power supplies are doing when there is no beam there as long as the main 295 kHz supply is putting out a magnetic field when beam is not expected (see Fig.~\ref{fig:power_supply}).

\subsection{Comment on Algorithm Error}

Given a 12-bit counter, ideally one would increment the counter by $2^{12}\times10~\text{nsec}/226~\text{nsec}=181.238932$, but since you can only increment by an integer, incrementing by 181 on every input clock cycle unavoidably introduces an error of 0.238932 to the counter on each clock cycle. 

However, since the phase accumulator is reset after every 113 input clock cycles (22, 23, 22, 23, 23 input clock cycles), the error only reaches a value of 27 before the reset. This is not enough to cause an early transition, and after 113 cycles the phase is reset and the cycle starts again from zero. Therefore, the approximation of the ideal counter value as 181 does not pose a problem in this design.

\subsection{Future Work}

Integration of this FPGA circuit with the real AC Dipole power supply and magnets still needs to be tested. Furthermore, the FPGA will rely on timing signals from upstream to tell it when the beam is coming after a pause so it knows to reset its phase, but that will come with a propagation delay, so the length of that propagation delay will need to be measured and accounted for. These are tasks that will be accomplished once the FPGA is connected to the real machines and can be tested in greater detail.


\ifboolexpr{bool{jacowbiblatex}}%
	{\printbibliography}%

\begin{thebibliography}{9} 

    \bibitem{TDR}
        Mu2e Collaboration, 
        ``Mu2e Technical Design Report'', 2015.
        \url{doi:10.48550/arXiv.1501.05241}


    \bibitem{extinction}
        J. Miller, 
        ``Beam Extinction Requirement for Mu2e'', Mu2e-doc-1175.
        \url{https://mu2e-docdb.fnal.gov}

	\bibitem{Werkema}
		S. Werkema,
		``Mu2e Proton Beam Longitudinal Structure'', Mu2e-doc-2771, 2019.
		\url{https://mu2e-docdb.fnal.gov}


    \bibitem{ac_dipole}
        E. Prebys and D. Still,
        ``Mu2e AC Dipole and Dipole Power Supply Specifications'',  Mu2e-doc-12113, 2017.\\
        \url{https://mu2e-docdb.fnal.gov}


    \bibitem{github}
        R. Hensley and S. Tripathi,
        ``AC Dipole FPGA Mechanics'',
        \url{https://gist.github.com/r-hensley/75a9a475550022aaeb189f50e06eb68d}
	
	\end{thebibliography}
	{%
	
	} 


%
%


\end{document}